%% file: paper.tex
\documentclass[letterpaper,11pt,leqno]{article}
\usepackage{paper}
\usepackage{adjustbox}
\usepackage{tabularx} 
\usepackage{makecell} 
\usepackage{siunitx} 
\pdfoutput=1

\hypersetup{pdftitle={Consumer Behavior under Benevolent Price Discrimination}}

\newcommand{\bib}{bibliography.bib}

\newcommand{\signstars}{* $p < 0.1$, ** $p < 0.05$, *** $p < 0.01$.}

\begin{document}

\title{Consumer Behavior under Benevolent Price Discrimination}

\author{Alexander Erlei, Mattheus Brenig, Nils Engelbrecht
%
\thanks{Corresponding author: Alexander Erlei, alexander.erlei@wiwi.uni-goettingen.de, ORCiD: 0000-0001-7322-2761. Chair for Economic Policy and SME Research, Georg-August-University Goettingen.\\
We gratefully acknowledge financial support by the Lower Saxony Ministry of Science and Culture under grant number ZN3492 within the Lower Saxony “Vorab“ of the Volkswagen Foundation and the Center for Digital Innovations (ZDIN)}}

\date{April 2024}   


\begin{titlepage}
\maketitle

Extensive research shows that consumers are generally averse to price discrimination. However, instruments of differential pricing can benefit consumer surplus and alleviate inequity through targeted price discounts. This paper examines how these outcome considerations influence consumer reactions to price discrimination. Six studies with 3951 participants show that a large share of consumers is willing to costly switch away from a store that introduces a discount for low-income consumers. This happens irrespective of whether income differences are due to luck or merit. While the price-discriminating store does attract some new high-income consumers, it cannot compensate the loss of existing consumers. Allowing for altruistic preferences by simulating a market mechanism increases costly support for price discounts, but does not alleviate consumer aversions. Finally, we provide evidence that warm glow drives costly support for price discounts.



\end{titlepage}

\section{Introduction}\label{s:introduction}
 
Technological progress, especially the advent of big data, has reduced the technological and informational constraints on sellers to differentiate prices among consumers \citep{calvano2020artificial,musolff2022algorithmic}. This development holds great economic potential for firms, as uniform prices prevent them from fully leveraging consumer valuations above the market price, while simultaneously pricing-out consumers with a lower willingness-to-pay (WTP). Importantly, it also implies potential benefits to consumer welfare and inequality. Those who are otherwise excluded from the market -- often due to lower financial means -- may be able to access previously closed markets under more personalized pricing, i.e. third-degree price discrimination. For example, \citet{dube2023personalized} present data from two field experiments showing that personalized pricing benefits the majority of consumers while also increasing seller profits. In turn, high-WTP consumers pay a substantially higher fee. \citet{10.1093/restud/rdad037} analyze data from international airline markets and find that using private information about passengers benefits overall welfare by increasing seller and leisure passenger surplus at the expense of business passengers (see also \citet{williams2022welfare}). 

Despite these straightforward incentives, most sellers still hesitate to deviate from uniform pricing strategies \citep{Fudenberg.2012,dellavigna2019uniform}. In a striking example, \citet{dube2023personalized} estimate that, in 2015, personalized pricing could have increased ZipRecruiter's expected profits by 86\% compared to their uniform price (and by 19\% compared to the theoretically optimal uniform price). One commonly evoked explanation for the lack of realized price discrimination is that sellers anticipate the risk of antagonizing consumers by violating their fairness perceptions \citep{Xia.2004,cox2004identify,fabiani2006firms,Haws.2006,Bolton.2010,Ferguson.2014}. Indeed, research has shown that profit-seeking through price discrimination is often inhibited by negative consumer reactions \citep{kahneman1986fairness,leibbrandt2020behavioral,allender2021price}. Recently, these sentiments have been echoed throughout many media outlets as well as reports from e.g. the US-American Council of Economic Advisors (CEA, 2015), the German Federal Data Ethics Committee (DEC, 2019) or the European Union (EU).\footnote{CEA: \url{https://obamawhitehouse.archives.gov/sites/default/files/whitehouse_files/docs/Big_Data_Report_Nonembargo_v2.pdf}; DEC: \url{https://www.bmi.bund.de/SharedDocs/downloads/DE/publikationen/themen/it-digitalpolitik/gutachten-datenethikkommission.html}; EU: \url{https://gdpr.eu/}} The latter even codified a right to complete price transparency into law, highlighting the consumer's crucial role for the adoption of price discrimination. Thus, in addition to technological constraints, firms also face ``behavioral constraints'' when deciding on their price setting strategy.

These observations open up an intriguing puzzle. On the one hand, third-degree price discrimination can increase overall economic welfare and simultaneously decrease inequity through reduced prices for low-income consumers. On the other hand, consumers social preferences appear to enforce uniform pricing and thereby prevent firms from opening up the market. This makes a straightforward argument for the need of an improved understanding about which factors determine consumer behavior under price discrimination, and the conditions that might alleviate seller constraints.

In this paper, we experimentally explore the behavioral mechanisms behind consumer reactions to differential pricing that favors low-income consumers who cannot afford the uniform price. We focus on \textit{benevolent price discrimination} (BPD), defined as a policy of differential pricing that is downward-bound and always benefits financially disadvantaged groups or individuals, thereby leading to a convergence of outcomes. Because sellers cannot increase prices, BPD improves joint welfare, while also creating a more equitable distribution of consumer gains. High-income consumers do not experience any absolute economic losses, but are \textit{relatively disadvantaged} by price discrimination. Thus, our main focus lies on situations where targeted price discounts allow previously excluded low-income consumers to participate in a market, without (negatively) affecting the welfare of other consumers. 

One prominent real-life example is the Supplemental Nutrition Assistance Program (SNAP) administered administered by the U.S. Department of Agriculture (USDA). In 2019, the department started a pilot program whereby consumers eligible for SNAP where allowed to use food stamps to buy groceries at participating online retailers, among them Walmart and Amazon. In addition to food-related items, Amazon offers some further benefits to SNAP-customers, such as an over 50\% discount for their subscription service Amazon Prime. They also do not try to obfuscate their participation, as eligible items are marked with an "SNAP EBT eligible" label. Hence, while there is a lot of evidence that consumers are generally averse to price discrimination, there appear to be exceptions. Indeed, there is a long history of sellers offering specific discounts to certain groups of the population, like students or the elderly. What it is exactly that drives the acceptance of such programs is not yet known, although several researchers have speculated that inequality concerns are likely to play an important role \citep{Rotemberg.2011,Wu.2012}. Identifying the specific mechanisms behind consumer choices from real-world data is complicated by the fact that consumers operate in a highly endogenous environment that usually allows for several distinct explanations. Even for a program like SNAP, eligibility hinges on several criteria such as income, work requirements, resource requirements, constrained housing expenditure, or immigrant status. Moreover, SNAP is publicly financed, participants only have access to a specific selection of products or services, and must exert personal effort during the application process. This makes it virtually impossible to parse which specific elements determine the programs apparent widespread acceptance. We therefore use an experimental approach to abstract from various cultural and context-dependent factors, thereby avoiding behavioral distortions induced by residual variables and isolating the effects of specific economic factors on consumer behavior.

Our conceptualization differentiates between (1) process-related fairness concerns and (2) social preferences over the outcome of an transaction, i.e. the realized economic gains and their distribution. There is unanimity in the pricing literature that consumers are generally averse towards the process of price discrimination, and thus often oppose different prices for the same good -- particularly if they are disadvantaged \citep{Campbell.1999, Haws.2006, maxwell2007price, leibbrandt2020behavioral}. Here, prices paid by other consumers, i.e. the consumer's peers, serve as an external reference point, whereby the consumer assesses if an offered price is fair \citep{thaler1985mental, janiszewski1999range, Jin.2014}.\footnote{Next to peer-induced fairness concerns, there are also distributional fairness concerns that relate to the seller's profit. Usually, these appear to be substantially less important as long as peer prices are not being obfuscated \citep{Ho.2009,allender2021price}.} Deviations from the reference point usually lead to averse reactions, including lower purchase satisfaction \citep{shor2006price}, intentions to spread negative word-of-mouth \citep{Ferguson.2014} and, most importantly, switching to a different seller or terminating the trade relationship \citep{Ho.2009, Ho.2014,anderson2008research, leibbrandt2020behavioral, allender2021price}. 

However, price discrimination that benefits low-income consumers, and especially BPD, also aligns positively with a number of pro-social concerns that have been thoroughly documented in the economic literature. One, there is ample evidence that people care about the equal distribution of economic outcomes \citep{fehr1993does, harrington2016relative}, particularly if existing income inequalities are seen as arbitrary \citep{rutstrom00, alesina05, durante14}. People tend to be inequity averse, and experience disutility when their outcome is different from other people's outcomes, or when they receive an unequal share of the total surplus \citep{Fehr.1999,bolton2000erc}. This effect is stronger for disadvantaged individuals, but has also been extensively documented for those with distributional advantages. Thus, as long as differential pricing leads to a convergence of consumer outcomes, and nobody incurs any additional losses, it may not only by accepted by consumers, but even be desirable.

Second, we argue that \textit{support} for BPD is consistent with impure altruism \citep{andreoni1990impure}. Altruistic people care about the economic outcomes of others, and will therefore incur utility losses if their actions harm others. In a competitive market environment, punishing sellers for introducing targeted price discounts discourages future BPD, which hurts low-income consumers in the long run by pricing them out of the market. Thus, altruistic preferences may alleviate negative consumer reactions towards BPD. In addition to that, research on impure altruism shows that many people derive non-monetary utility through "warm glow" when helping others. For example, when donating to a charity, people experience private benefits that are not motivated by the gains of the recipients, but the contribution as such. People feel good about donating to charities, even if they provide no tangible benefits. Beyond the economics of giving \citep[see also][]{ottoni2017people,korenok2013impure}, private provision of public goods through altruism or warm-glow has also been documented in cooperative \citep{andreoni1995cooperation,palfrey1997anomalous,goeree2002private} and labor-market like gift-exchange environments \citep{cox2004identify,charness2002altruism}. We argue that switching towards a price-discriminating store that offers goods to low-income consumers at reduced prices might evoke similar non-monetary rewards, thereby increasing the private provision of attainable prices for low-income individuals. This implies costly consumer support for sellers who introduce BPD. 

To the best of our knowledge, we are the first to systematically analyze consumer reactions towards price discounts that are exclusively targeted at low-income consumers who were previously excluded from the market. We are also the first ones to quantify the effects of consumer altruism and warm glow preferences on sellers price setting constraints. 

\subsection*{Study Overview}

We conduct a series of pre-registered, incentivized, context-neutral and controlled experimental studies using Amazon Mechanical Turk (MTurk) and CloudResearch. Sellers can offer a price discount for previously excluded low-income consumers, enabling them to purchase in their store and thereby leading to a convergence in consumer outcomes. Price discrimination is transparent, and high-income consumers can either costly switch to a second seller who does (not) offer the discount, choose not to buy at all, or maximize their income by staying with the same seller across all rounds. By quantifying both, behavioral shifts away and towards the price discriminating seller, we capture bidirectional consumer migration patterns that have so far been largely neglected by the experimental literature. This allows us to make qualitative judgments about the net-effect of BPD, rather than focusing solely on the behavioral constraints elicited by current consumers.

The first three studies establish consumer's baseline behavior under BPD and analyze the impact of inequity concerns. Our results show widespread aversions towards BPD. Across the three experiments, 30\% -- 40\% of consumers exhibited strictly BPD-averse purchasing patterns, with little to no effect evidence that an equalization of outcomes mitigates negative reactions. Controlling for reciprocity concerns by keeping sellers out of the price-setting process does not change the results, and high-income consumers appear indifferent to the cause of income inequality. That is, irrespective of whether the initial consumer incomes are allocated randomly or via a subject's performance in a real-effort task, aversion remains high. Costly support for the BPD-seller is virtually non-existent. 

Study 4 introduces a market mechanism that gives high-income consumers some agency over future prices, i.e. future low-income consumer outcomes. This allows for altruism, and possibly induces feelings of warm glow. We find marginal evidence that altruism may slightly alleviate consumer constraints towards BPD when income differences are generated randomly. Generally, aversions persists. Contrary to studies 1 -- 3, a sizeable number of consumers is willing to costly \textit{support} targeted price discounts by switching towards the price-discriminating seller when their purchases influence the store's future pricing strategy. Allowing for altruism doubles the share of consumers who exhibit strictly pro-pd behavior, and more than triples the number of goods sold by the seller. However, consumer aversions still substantially outweigh consumer support.

Study 5 systematically varies switching costs for both paradigms. In line with warm glow effects, we expect higher switching costs to increase costly consumer support for BPD, and reduce the share of consumers switching away from the price discriminating seller. The experimental data confirms these predictions.

Finally, we provide two robustness checks that are detailed in the online appendix.\footnote{The complete data set, additional information and this project's code can be accessed under the following repository: \url{https://osf.io/sztnh/?view_only=fc170dcbfeda431e994b23ed095d8b4b}} First, using qualitative data, we confirm the validity of our behavioral interpretations, showing that a large share of consumers reports to costly switch away from the price discriminating store due to being unfairly passed over for the price discount. Second, we conduct a sixth large study that varies low-income consumers willingness-to-pay for the product, showing that consumer aversions and support are independent of purchase-level inequity attitudes.

\section{Experimental Design and Theory}\label{s:design}

Because the basic experimental setup was the same across all studies, we will describe it here in more detail and refer to this description in the remainder of the paper. 

\subsection*{Experimental Design}

We ran purchasing experiments where consumers had the option to buy a homogeneous good at one of two stores, each run by another participant acting as store manager. We measure preferences for or against price discrimination by a consumer's store choice. We included a post-experimental questionnaire measuring participants' fairness perceptions, attitudes towards price discrimination as well as a questionnaire on social comparison \citep{gibbons99}. Summary statistics can be found in the online appendix. All experiments in this paper were conducted online using MTurk, CloudResearch and oTree \citep{chen16}. Participants enrolled on their own accord and were randomly assigned to one treatment and one role. After the instructions, participants had two attempts to answer five comprehension questions correctly. Those who failed both attempts were excluded from participating. We used ``Coins'' as our experimental currency. Coins were later converted into dollars, where 10 coins equaled 3 cents. Additionally, subjects playing the role of ``consumer'' received a fixed payment of \$1.20 for completing the survey. There were two main experimental paradigms to analyze consumer behavior: \textit{One-Shot} and \textit{Repeated}.

\subsubsection*{One-Shot}

Upon arrival, participants were randomly assigned the role of either a \textit{high-income} or a \textit{low-income} consumer. High-income (low-income) consumers learned that they would receive an endowment of 100 (50) Coins. Coins could be used to buy a good at one of two stores (A, B). In both stores, the good returned a value of 150 Coins to a consumer's final payoff. Thus, consumers were incentivized to always buy the good. The two stores sold the same good and were identical except for one thing: one store offered all consumers the good for a price of 100 Coins. The other store offered low-income consumers the good for a discounted price of 50 Coins and charged high-income consumers the regular price of 100 Coins. Consumers learned that each store was run by a manager, who had decided on the pricing strategy of the store beforehand, and that managers earned money for each good sold in their store. This ensured that consumer purchasing choices were meaningful. 

Participants decided at which of the two stores they wanted to purchase the good. High-income consumers could observe that one store offered the good at a price of 50 Coins and learned with a click that they were not eligible for that discount. However, in Studies 1 and 2, no explanation for the different prices was given. We randomized the position (left, right) and name (A, B) of the store introducing price discrimination. Thus, if price discrimination had no impact, we would expect consumers to be equally distributed across both stores.

\subsubsection*{Repeated}
Compared to the one-shot design, consumers completed four purchasing rounds. We did not change the basic parameters of the experiment. High-income consumers received 100 Coins each round and could use them to purchase the good in either store A or store B. Low-income consumers were not able to purchase a good in the first two rounds and had to rely on an outside option that was a simple multiplier of one and always available to all consumers. Endowments could not be transferred from one round to another. Thus, we can rule out that consumers may respond to a switch in sellers pricing strategies by delaying their purchases in anticipation of future discounts \citep{coase1972durability}.

To make switching meaningful, we introduced monetary costs for switching between stores in two consecutive rounds. Consumers always started a round in the store they chose the previous round. For example, if a consumer purchased the good in store A in the first round, they started the second round in store A. If the consumer then decided to switch and purchase the good in store B, they had to pay a fee of 10 Coins, i.e.\,10\,\% of their endowment per round, after the purchase. Thus, consumers were monetarily disincentivized to switch.

In the first two purchasing rounds, both stores offered the good for the same price. After the second round, consumers learned that one of the two store managers changed their pricing strategy and would offer low-income consumers the good for a price of 50 Coins in the remaining two rounds. Depending on treatment, this was either (i) the manager of the store consumers had purchased in during the second round (\textit{Avoid}) or (ii) the manager of the store consumers had \textit{not} purchased in during the second round (\textit{Approach}). Price discrimination was transparent to all consumers.

This setup imposes stricter conditions on observing consumer preferences against price discounts. Subjects who switch away from (or towards) the price discriminating store are willing to substantially reduce their own payoff. Second, it allows us to analyze benevolent price discrimination, where consumers who already participate in the market do not experience a price increase and only low-income consumers who priorly could not purchase any goods benefit from lower prices. Whereas in the one-shot design, consumers might perceive the lower price to be at their own costs, the within-subject design establishes the reference price of 100 Coins over two rounds. Importantly, low-income consumers miss out on substantial payoff over the first two rounds, and the price discount in rounds three and four serves to mitigate, but not fully eliminate unequal outcomes. 

\subsubsection*{Manager}

In each treatment, two participants were assigned the role of a store manager. Before consumers made their choices, managers were free to decide whether they wanted to implement a discount for low-income consumers in their store. In the \textit{Repeated} treatments, price discounts could only be introduced after the second round. Managers had full information about the experimental setup. We gathered observations until we had one manager who decided for and one who decided against price discrimination for each experimental treatment. Hence, no deception was involved. Managers were rewarded with a base reward of \$0.50 and earned 1 Cent for each good sold in their store. 

\subsection*{Theory}

To illustrate how the different variables described above might affect purchasing behavior under BPD, we introduce a short stylized theoretical section. The model concentrates on high-income consumer's switching decision in the \textit{Repeated} paradigm after the introduction of price discounts for low-income consumers. High income consumers can use their endowment $e_{hc}$ to buy a good at their current store ($s=0$) that returns $w$ for price $p_{hc} = e_{hc}$. Alternatively, the high-income consumer can switch to a second store ($s=1$) and buy the same good for the same price, but pay additional switching costs $c$. In \textit{Avoid}, the current store introduces price discrimination, in \textit{Approach}, the second store reduces prices for low-income consumers ($p_{lc}$). Thus, $s=1$ measures preferences against price discrimination in \textit{Avoid}, and preferences for price discrimination in \textit{Approach}. For the very basic case where high-income consumers receive no contextual information and observe price discounts, switching behavior for \textit{Avoid} might be described by a simple reference point model \citep[see][]{leibbrandt2020behavioral}:
\begin{equation}
    U_{i}^{avoid}(w,p_{hc},p_{lc}) =
    \begin{cases}
        w - \gamma_{i}(p_{hc}-p_{lc}) & \text{if } s = 0 \\
        w - c & \text{if } s = 1
    \end{cases}
\end{equation}
Here, $\gamma_{i} \geq 0$ captures an individual's aversion towards paying a higher price than other consumers, i.e. disutility induced by procedural price unfairness ($p_{hc} > p_{lc}$).\footnote{We do not explicitly analyze social preferences over seller surplus. Our experimental setup minimizes seller information, and focuses on peer-comparison between consumers. The reference point model also does not exclude disutility due to perceived seller gains. In prior work on bargaining where people negotiate with a seller in the presence of peers, fairness concerns towards peers were substantially stronger than towards the seller \citep{Ho.2009}.} For \textit{Approach}, utility-maximizing high-income consumers (almost) never switch because they cannot meaningfully evaluate the price changes beyond the reference point effect. In that case, not switching is clearly the dominant choice:

\begin{equation}
    U_{i}^{approach}(w,p_{hc},p_{lc}) =
    \begin{cases}
        w  & \text{if } s = 0 \\
        w - c - \gamma_{i}(p_{hc}-p_{lc})  & \text{if } s = 1
    \end{cases}
\end{equation}

\subsubsection*{Inequity Aversion and (Impure) Altruism}
When high-income consumers learn about the benevolent nature of price discrimination, switching away becomes less attractive. First, the existence of low-income consumers induces inequity considerations. Specifically, high-income consumers in \textit{Avoid} who change stores might experience disutility due to advantageous inequity concerns \citep{Fehr.1999}\footnote{High-income consumers can never earn less than low-income consumers. Hence, there are no disadvantageous inequity concerns.}:
\begin{equation}
    U_{i}^{avoid} =
    \begin{cases}
        w - \gamma_{i}(p_{hc}-p_{lc}) & \text{if }  s = 0  \\
        w - c   -  \beta_{i}\max{\{0,w - e_{lc}\}} & \text{if } s = 1 
    \end{cases}
    \label{eq3}
    \end{equation}  
The degree of aversion from being distributionally ahead is captured by $\beta_{i} \geq 0$. For example, a consumer who perceives their high-income status to be a result of random luck might exhibit a higher $\beta$ compared to those who earned larger endowments through hard or productive work. Staying in the price-discriminating store alleviates inequity considerations, as both consumer types earn the same amount of money by purchasing the same good.

Conversely, BPD in \textit{Approach} causes new opportunity costs for staying in the store without price discounts because high-income consumers receive the option to decrease inequities by switching towards the BPD-store:

\begin{equation}
    U_{i}^{approach} =
    \begin{cases}
        w - \beta_{i}\max{\{0,w - e_{lc}\}} & \text{if }  s = 0  \\
        w - c - \gamma_{i}(p_{hc}-p_{lc}) & \text{if }  s = 1
    \end{cases}
\end{equation}

Second, in addition to the alleviation of inequity concerns, high-income consumers who purchase at the BPD-store and thereby support higher economic returns for low-income consumers might derive \textit{warm glow} and \textit{altruism} utility \citep{andreoni1990impure}. Warm glow refers to the satisfaction an individual experiences through the act of giving to others. It might be the primary explanation for giving behavior -- such as charitable donations -- in large-scale economies \citep{ribar2002altruistic,luccasen2017warm}. In our setup, warm glow only functions in the \textit{Approach} paradigm. That is because high-income consumers who approach the BPD-store essentially pay a cost, $c$, to support a purchasing option that increases low-income consumer income.\footnote{Since warm glow functions independent of the recipient's output, it is not even necessary to assume that purchasing behavior by high-income consumers actually influences market supply. However, because consumers in the current setup are second movers, the effect might still depend on the existence of some future impact that allows consumers to frame their switching decision as a "good deed".} It is thus closely analogous to the private provision of a re-distributive, charitable mechanism. High-income consumers choose between consuming $w$ or $w - c$, where the switching costs determine (and constrain) the amount of quasi-giving. In return, they receive additional utility $k_{i}(c)$. Furthermore, consumers in both paradigms might exhibit altruistic preferences towards low-income consumers and therefore derive utility from BPD because it increases the latter's consumption from $e_{lc}$ to $w$. We define $\pi = w - e_{lc}$. Following impure altruism, the additional utility from switching towards the BPD-store $k_{i}(c,w-e_{lc})$ then depends on both warm glow and altruism:  

\begin{equation}
    U_{i}^{approach} =
    \begin{cases}
        w - \beta_{i}\max{\{0,w - e_{lc}\}}
        & \text{if } s = 0 \\
        w - c - \gamma_{i}(p_{hc}-p_{lc}) + k_{i}(c,\pi)  
        & \text{if } s = 1
    \end{cases}
    \label{eq5}
\end{equation}

Here, $k_{i}(c,\pi)$ is a quasi-concave, strictly increasing utility function for individual $i$. Note that $k_{i}(c,\pi)$ only holds if consumers' switching behavior influences the pricing strategy of the BPD-store (Study 4 \& 5). Otherwise (Study 3), additional utility only depends on giving ($k_{i}(c))$. Finally, altruistic preferences can increase the utility costs of avoidance behavior when consumer choices influence future price-setting. Switching away from the seller with price discounts reduces the probability of lower prices for low-income consumers in the future and thereby decreases their potential consumption level:

\begin{equation}
    U_{i}^{avoid} =
    \begin{cases}
        w - \gamma_{i}(p_{hc}-p_{lc}) & \text{if } s = 0 \\
        w - c   -  \beta_{i}\max{\{0,w - e_{lc}\}} - k_{i}(\pi) & \text{if } s = 1
    \end{cases}
    \label{eq6}
\end{equation}

\begin{table}[t!]
\caption{High-income consumer utility functions by study and switching paradigm.}
\begin{adjustbox}{max width=\textwidth}
\begin{tabular}{llllll}
\toprule
             & \multicolumn{2}{c}{$U_{i}^{avoid}$} & & \multicolumn{2}{c}{$U_{i}^{approach}$} \\ \cmidrule{2-3}\cmidrule{5-6}
             & \multicolumn{1}{c}{Stay}       & \multicolumn{1}{c}{Switch}       && \multicolumn{1}{c}{Stay}         & \multicolumn{1}{c}{Switch}        \\ 
             \cmidrule{2-3}\cmidrule{5-6}
Study 1 \& 2 &    $w - \gamma_{i}(p_{hc}-p_{lc})$         & $w - c$            & &  $w$            &    $w -c - \gamma_{i}(p_{hc}-p_{lc}) $           \\
Study 3 &  $w - \gamma_{i}(p_{hc}-p_{lc})$      &     $w - c   -  \beta_{i}\max{\{0,w - e_{lc}\}} $                  & & $w - \beta_{i}\max{\{0,w - e_{lc}\}}$              &     $ w - c - \gamma_{i}(p_{hc}-p_{lc}) + k_{i}(c)  $          \\
Study 4, 5 \& 6 &  $w - \gamma_{i}(p_{hc}-p_{lc})$      & $w - c   -  \beta_{i}\max{\{0,w - e_{lc}\}} - k_{i}(\pi)$             & & $ w - \beta_{i}\max{\{0,w - e_{lc}\}}$              &     $ w - c - \gamma_{i}(p_{hc}-p_{lc}) + k_{i}(c,\pi) $          \\ 
\bottomrule
\end{tabular}
\end{adjustbox}
\note{}
\label{tab:theory}
\end{table}

As Table \ref{tab:theory} shows, our studies gradually build on these stylized intuitions, allowing for increasingly more social preferences to determine switching behavior. 

\section{Studies 1 -- 3: Reference Prices and Inequity Aversion}\label{s1-3}

We first present three studies that quantify baseline consumer behavior under benevolent price discrimination and introduce the possibility of inequity concerns. In the first two, subjects do not receive information about the beneficiaries of BPD, and cannot infer that the discounts lead to more equal outcomes. This precludes social considerations or preferences from affecting purchasing choices and provides a baseline for consumer behavior towards relatively disadvantageous price discounts. It is also a relevant approximation of many real-world contexts, where it is virtually impossible or too costly for consumers to verify the specific circumstances behind differential pricing. 

The third study extends that design by two crucial features. First, consumers observe income as the variable of discrimination. Second, we introduce different causes of income inequality. Endowments are either allocated randomly (\textit{Random}) or based on the performance (\textit{Effort}) in a real-effort slider task \citep{gill12}.\footnote{Consumers receive either the high or the low endowment based on a fixed performance threshold of 31 sliders over three rounds of 60 seconds each. Here, we follow a pretest with 100 participants and select the cut-off at the 10th percentile.} Consumers in \textit{Effort} know that their performance influences their endowment, but are not informed about the exact functional relationship, and thus do not know the specific endowment levels of other consumers. 

Compared to studies 1 -- 2, high-income consumers know that (i) low-income consumers exist, (ii) low-income consumers cannot afford to buy a good for the initial uniform price, (iii) only low-income consumers benefit from price discrimination and (iv) high- and low-income consumers can afford the same amount of goods under price discrimination. Hence, Study 3 reveals the benevolent nature of price discounts to high-income consumers. Importantly, this also implies that low-income consumers can never earn more than, or even the same as high-income consumers. This rules out disadvantageous inequity as a motivation for high-income consumers. 

If consumer behavior towards BPD is determined by inequity considerations, we would expect fewer subjects to costly switch away from the BPD-seller in Study 3. Furthermore, high-income consumers in \textit{Effort} may exhibit stronger behavioral aversions than those in \textit{Random} because BPD undermines a merit-based advantage, which decreases the inequity parameter $\beta$. For \textit{Approach}, we expected some consumers to costly support the price discriminating store due to warm glow effects. However, such an effect depends on the consumers ability to frame their purchasing choice as a "good deed", which may require the \textit{possibility} of having some tangible impact on other peoples utility. In that case, costly support would not differ between studies. 

\subsection*{Experimental Design}
Studies 1 and 2 follow the basic procedure outlined above. In Study 1, participants are randomly assigned to \textit{One-Shot}, \textit{Repeated-Avoid} or \textit{Repeated-Approach}. We gathered data until we had 100 high-income consumers in each treatment and set a fixed probability of 10\% to become a low-income consumer.

Study 2 focuses on \textit{Repeated-Avoid} to control for reciprocity towards the seller as a main driver of consumer behavior. Without any information about either the goal, reasons or beneficiaries of price discrimination, consumers are free to interpret the manager's actions in a variety of manners. Following attribution theory, people exhibit strong tendencies to infer causes and assign responsibilities or intentions when they perceive some outcome to be unfair \citep{heider1982psychology,blount1995social,falk2008testing}. This may induce high-income consumers to punish managers because they want to negatively reciprocate the act of being intentionally passed over for a discount. To address this, we eliminate any intentionality behind price discrimination by keeping managers out of the price-setting process. Instead, consumers learn that each manager decided to install a pricing algorithm, without the ability to predict changes. For a detailed description of the study, please refer to the online appendix. Here, we only focus on the results, to substantiate the importance of peer-comparison for consumer behavior. We run two treatments (\textit{Intention} vs. \textit{No Intention}) with 200 observations each.

Finally, Study 3 consists of six treatments within a 2 (\textit{Random} vs. \textit{Effort}) x 3 (\textit{One-Shot} vs. \textit{Repeated-Avoid} vs. \textit{Repeated-Approach}) between-subject design. We gathered observations until we had 100 high-income consumers in each treatment with a fixed probability of 10\% to become a low-income consumer. The basic procedure mirrors Study 1. In \textit{Random}, we add one page where participants learn about the random endowment mechanism. In \textit{Effort}, participants first complete three rounds of a slider-based effort task. After being informed about their endowment, we ask whether it would be fair if every consumer received the same endowment, and whether it is fair that (1) endowments are allocated randomly or (2) better performances in the slider task are rewarded with a higher endowment. Consumers then proceeded with the original instructions. Those who completed the effort task but failed to answer the comprehension questions were paid the base reward of \$1.20.

\subsection*{Study 1 \& 2}
We always exclude all observations where a subject chose the outside option in the second round. This leaves us with 96 independent observations in \textit{Avoid}, 97 independent observations in \textit{Approach} and 100 in \textit{One-Shot}. See Table \ref{tab:study_overview} for an overview of participant-samples and treatments across all studies in this article.

\input{tab_overview}

\subsubsection*{One-Shot}

From now on, we will refer to high-income consumers as ``consumers''. Low-income consumers are irrelevant for our analysis and only serve to avoid deception. As hypothesized, a large majority of consumers prefers the non-price-discriminating store. Without any explanation for the observed price discrimination, 86\,\% of consumers purchase the good in the store that charged all consumers the same price, 9\,\% purchase in the store with price discounts and 5\,\% choose the outside option.
A one sample $t$-test also confirms consumers to significantly prefer the non-price-discriminating store ($t = 14.25, p < 0.001$). The name $(\tilde{\chi}^2 = 1.33, p = 0.25)$ and position  $(\tilde{\chi}^2 = 1.66, p = 0.20)$ of the store introducing price discrimination do not have a significant effect on store choice.

\subsubsection*{Repeated-Avoid}

\begin{figure}[htbp]
\begin{center}
\begin{minipage}{0.9\linewidth}
\includegraphics[width=1\textwidth]{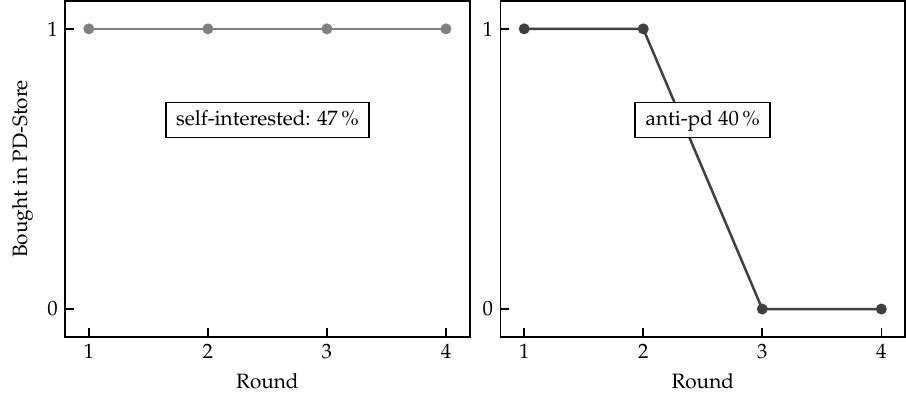}
\caption{The two most common behavioral patterns in \textit{Avoid}}
\label{fig: s1_within_avoid}
\note{The two most common purchase patterns were ``self-interested'' and ``anti-pd''. Consumers classified as self-interested never switched to the other store. Consumers classified as anti-pd stayed within the same store for the first two rounds and then switched after their store introduced price discrimination.}
\end{minipage}
\end{center}
\end{figure}

In the first two rounds, both stores offer the good for the same price. After the second round, consumers learn that the store they chose in the second round offers some consumers lower prices. Figure \ref{fig: s1_within_avoid} illustrates the two most common behavioral patterns in the \textit{Avoid} treatment. 

We find that 47\,\% of consumers never switch and maximize their payoff. We classify those consumer as ``self-interested''. Surprisingly, almost as many consumers decide to costly switch away from a benevolently price-discriminating store. We classify those consumers who stay in the same store for the first two rounds, switch after the second round, and stay in the other store for the third and fourth round as ``anti-pd''. Note that we can largely rule out distributional motives towards the managers' income as a reason for consumer switching for two reasons. One, we did not tell participants how many rounds they would play. Thus, consumers who wanted both managers to earn the same would have already switched in the second round. A few consumers exhibited these egalitarian preferences. Two, we asked subjects after the experiment \textit{Who do you think made more profit throughout this task?} Here, the majority of ``anti-pd'' (79\%) and ``self-interested'' (86\%) subjects stated that the manager they switched towards (or stayed in) made more or equal profits. This result replicates throughout the studies of this article. Subjects do not choose the manager they think to be worse off.\footnote{Generally, the most popular answer was that the manager subjects purchased in after the second round also made the largest profits. Additionally, text data from an open-ended question in the questionnaire of later studies also strongly confirms the validity of our interpretation. Almost no consumer ever refers to either manager's income.}

 40\,\% of consumers exhibit the conservative ``anti-pd'' pattern. We run a paired $t$-test to test whether consumers are less likely to purchase in the price discriminating store after it has introduced price discrimination, i.e. in round three and four. On average, consumers purchase 1.91 out of 2 goods in round one and two in the price-discriminating store, and only 1.04 out of 2 in rounds three and four $(t = -8.64,\, p < 0.001)$. 
 
 Throughout this paper, we will use a random effects panel logistic regression model with clustered standard errors to confirm and expand on our findings. In accordance with the results above, we report a significant and large negative effect of the BPD-dummy on a subject's probability to purchase in the price-discriminating store (see Tables \ref{tab:reg_avoid} and \ref{tab:reg_approach} in the appendix).

\subsubsection*{Study 2: Intentions}
 
 We replicate our results from Study 1 when controlling for manager intentions by keeping them out of the price-setting process. In \textit{Intention} (\textit{No Intention}), 37\,\% (30\,\%) of consumers are willing to costly switch away from a benevolently price-discriminating store, while 45\,\% (49\,\%) act monetarily self-interested. The average number of goods sold by the price-discriminating store decreases significantly with the introduction of price discrimination in round three and forth in both \textit{Intention} ($t = -13.23, p < 0.001)$ and \textit{No Intention} ($t = -11.02, p < 0.001$). There is moderate evidence that controlling for manager intentions slightly alleviates consumer aversions, as the regression model reveals a significant interaction between the introduction of BPD and \textit{Intention} (see Table \ref{tab:reg_avoid}), suggesting that subjects are less likely to purchase in the price-discriminating store after the introduction of BPD when the decision is made by a store manager instead of a pricing algorithm. However, a $t$-test comparing the average number of goods sold by the price-discriminating store during rounds three and four in \textit{Intention} (0.98) and \textit{No Intention} (1.09) shows no significant difference ($t = 1.1, p = 0.270$).We take these results as evidence that consumer aversion towards downward price-discrimination is not primarily driven by negatively reciprocal actions following an ascription of manager intentions.

\subsubsection*{Repeated-Approach}

In \textit{Approach}, the store not chosen by the consumer in the second round introduces price discrimination. This setup was designed to capture positive reactions towards BPD. As predicted, 80\,\% of consumers follow the self-interested pattern, whereas only 4\,\% switch to the price-discriminating store. A paired $t$-test shows no significant differences in the likelihood that a consumer purchased the good in the price-discriminating store between rounds 1 and 2 versus 3 and 4. On average, consumers purchase 0.07 out of 2 goods in round one and two in the price-discriminating store, and 0.12 out of 2 after the introduction of price discrimination in rounds three and four $(t = 0.96,\, p = 0.339)$. 

\subsection*{Study 3}
\subsubsection*{One-Shot}
In \textit{Effort}, consumers find a hypothetically equal distribution of endowments significantly more unfair than in \textit{Random} (\textit{Effort}: 2.52, \textit{Random}: 6.42; $t = 18.31,\, p p < 0.001)$. They also find the mechanism by which differences in endowments are achieved significantly fairer (\textit{Effort}: 6.37, \textit{Random}: 4.22; $t = -9.34,\, p < 0.001)$. This affirms that subjects in \textit{Effort} perceive differences in income and thereby different purchasing abilities as more fair, presumably because they are rooted in merit. Regarding store choices, consumers in the one-shot experiment still exhibit strong aversions against price discrimination, albeit weaker than in Study 1. In \textit{Random}, almost 67\,\% of consumers choose to purchase in the non-price-discriminating store. In \textit{Effort}, the share is around 72\,\%. The difference is not significant $(t = 1.46, p = 0.146)$.  

\subsubsection*{Repeated-Avoid}
\begin{figure}[t]
    \centering
    \includegraphics[width=\textwidth]{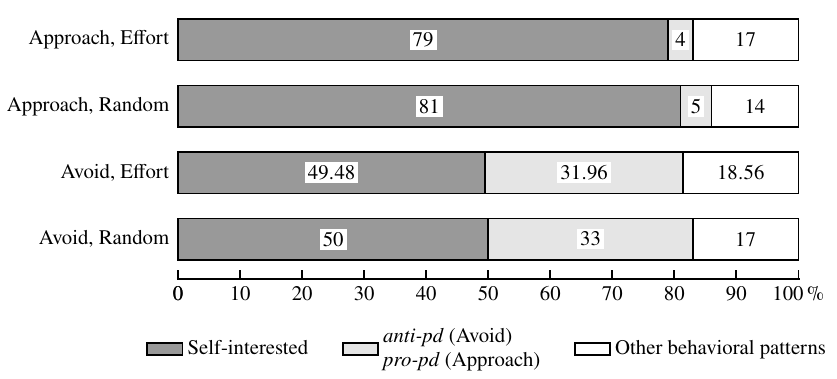}
    \caption{Behavioral patterns across the four conditions from Study 3.}
    \label{figure: s2_within_avoid}
\end{figure}

Results for the manipulation check are in line with the ones from \textit{One-Shot} and confirm the success of our intervention.
In \textit{Avoid}, we largely replicate the results from Study 1 for both \textit{Effort} and \textit{Random} (see Figure \ref{figure: s2_within_avoid}). 

Surprisingly, we find no differences between price discrimination over merit-based compared to arbitrary income inequality. In both treatments, one third of consumers is willing to costly switch to a non-price-discriminating store. Note that this is with the information that only low-income consumers who are otherwise not able to participate in the market benefit from price discrimination, while nobody incurs any income losses. Consumers purchase significantly less goods in the price discriminating store after the introduction of price discrimination (\textit{Effort}: 1.11 vs. 1.86, $t = -7.61, p < 0.001$; \textit{Random}: 1.11 vs. 1.88, $t = -8.22, p < 0.001$). There is no significant treatment difference $(t = -0.10, p = 0.923)$ and a random effects logistic regression reveals no significant interaction between \textit{Effort} and the introduction of BPD over the whole sample (see Table \ref{tab:reg_avoid}).

\subsubsection*{Repeated-Approach}
In accordance with Study 1, almost no consumer is willing to costly support a benevolently price discriminating store. Instead, the majority of consumers exhibits self-interested behavior patterns (\textit{Effort}: 79\,\%, \textit{Random}: 81\,\%). Only four and five percent respectively follow the pro-pd pattern. Paired $t$-tests suggest that consumers do not significantly change their behavior once the other store introduces price discrimination (\textit{Effort}: 0.17 vs. 0.11, $t = 1.10, p = 0.275$; \textit{Random}: 0.15 vs. 0.11, $t = 0.68, p = 0.495$).

\subsection*{Discussion}
A large share of consumers rejects relatively disadvantageous price-discrimination and is willing to pay money in order to switch towards a seller with uniform pricing. This pattern holds for BPD where consumers are aware that price discounts only benefit low-income consumers who are otherwise excluded from the market. Hence, the equalization of consumer outcomes does not appear to be a sufficient condition for sellers to introduce targeted price discounts. In contrast to many other documented scenarios in the context of economic inequality and re-distributive preferences, consumers appear to disregard the cause of income inequality. One possible explanation might simply be that procedural fairness judgments, i.e. charging different prices for the same good, are much more important in determining consumer behavior than outcome-related reasoning. This would also be in line with research showing that negative reciprocity is both more common and more intense than positive reciprocity, at least in one-shot settings \citep{baumeister2001bad,offerman2002hurting,al2009experimental}. 

An alternative explanation is that the top-down approach by which sellers impose differential pricing reduces both the moral responsibility and agency of consumers in causing more equal outcomes. Because in our setup the ``future market'' is unaffected by consumer decisions, and consumers do not decide whether low-income consumers receive discounts, they might exhibit patterns that are different from e.g. voting for re-distributive policies. In the economics literature, this has been referred to as responsibility-alleviation \citep{charness2000responsibility}. Put another way, decoupling high-income consumer choices from low-income consumer outcomes isolates for warm-glow-altruism. Altruists who care about the outcome of the ``public good'' have no reason to switch toward the BPD-store, or incur the personal costs of staying in a store that advantages others. To test these explanations, Study 4 extends our design by a second period that simulates consumer agency through a market environment.

 \section{Study 4: Altruism}\label{s4}

We examine whether consumer aversions towards BPD generalize to a market-analogous situation where consumers' purchasing decisions affect a store's future pricing strategy. By endowing current consumers with agency over the price-setting process for a later period, we induce externalities on their purchasing decisions and thus agency regarding future consumer outcomes. Consumers are responsible for future overall welfare as well as the potential equalization of outcomes induced by BPD. In so far as our subjects exhibit altruistic preferences, we would expect less switching in the \textit{Avoid} paradigm and more switching in \textit{Approach}. Those who still switch away from the price discriminating store essentially pay to decrease seller and future low-income consumer profits as well as to prevent a more equal outcome distribution.

\subsection*{Experimental Design}
We implemented two (\textit{Avoid} vs. \textit{Approach}) 2x2 mixed factorial designs manipulating the source of income inequality (\textit{Random} vs. \textit{Effort}) as well as the agency of consumers regarding the price-discriminating store's pricing strategy in a future task (\textit{Agency} vs. \textit{No agency}). The \textit{No Agency} treatments are equivalent to the corresponding \textit{Avoid} conditions in Study 3. For the agency treatments, we add one screen after the store's introduction of price discrimination. On this screen, subjects are informed that the discriminating store's sales in the remaining two rounds will determine its price-setting in a future HIT on MTurk. If sales exceeded a certain threshold, a rule would automatically implement the lower prices for new low-income consumers in the last two rounds of a future setup-equivalent task. Otherwise, future low-income consumers would face the uniform price, and thus be unable to purchase the good at all. We thereby foreclose any considerations regarding differing manager income levels or reverse outcome inequalities in the future task.\footnote{Specifying that low-income consumers would receive the discount only for two rounds ensures that it is impossible for them to earn more money in total than regular consumers.}

\subsection*{Results Avoid}

\textbf{Manipulation Check.} To affirm that subjects understand our manipulation, we asked them to choose the correct statement about the future impact of their purchase decisions out of four options. In total, 76\,\% of subjects answered the questions correctly. Since there could be multiple reasons why participants might answer the question incorrectly, we do not drop observations based on the manipulation check. Instead, we will use the results from the sub-sample of subjects who correctly answered the question as a robustness check.

\begin{figure}[t]
    \centering
    \includegraphics[width=\textwidth]{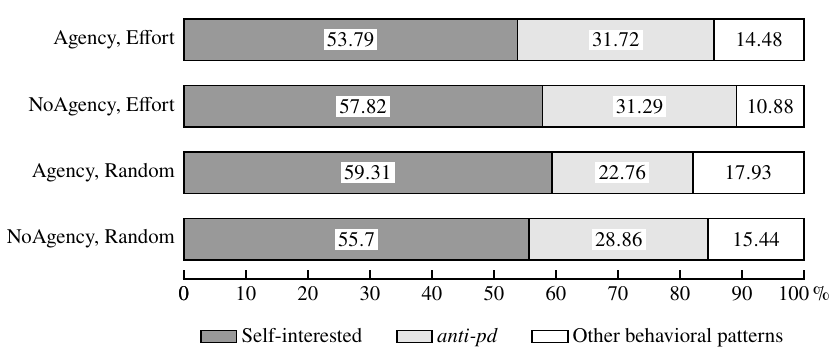}
    \caption{Behavioral patterns across the four conditions from Study 4, \emph{Avoid}}
    \label{fig:overview_4a}
\end{figure}

\noindent
\textbf{Switching Behavior.} In line with Study 3, 58\,\% (56\,\%) of subjects in the \textit{No Agency} treatments of \textit{Effort} (\textit{Random}) decide to stay in the same store for the whole experiment. Similarly, 31\,\% and 29\,\% respectively exhibit strictly anti price-discrimination behavioral patterns by switching after the introduction of discounted prices. Endowing subjects with agency regarding future consumer welfare does not appear to fundamentally shift behavior (see Figure \ref{fig:overview_4a}).

In \textit{Agency-Effort}, 32\,\% are categorized as \textit{anti-pd}. In \textit{Random}, that share drops to 23\,\%. Contrary to the \textit{No Agency} treatments, we document a difference in the drop of sales conditional on the source of endowment differences. In \textit{Random}, the store introducing BPD experiences a smaller decline in demand (\textit{Effort}: -0.72, \textit{Random}: -0.54; $t = -1.56, p = 0.06$), which is accentuated when restricting the sample to subjects who correctly answered the control question ($t = -1.89, p = 0.03$). The regression model replicates this effect across the whole sample through a significant interaction between the \textit{Effort}-Indicator and the BPD-Indicator.
Hence, there is moderate evidence that under agency, the source of endowment differences becomes more meaningful and consumers who randomly receive a higher income tend to switch less. However, differences are small and only marginally significant, depending on the sample. The overall behavioral patterns remain largely unchanged.

\subsection*{Results Approach}

\begin{figure}[t]
    \centering
    \includegraphics[width=\textwidth]{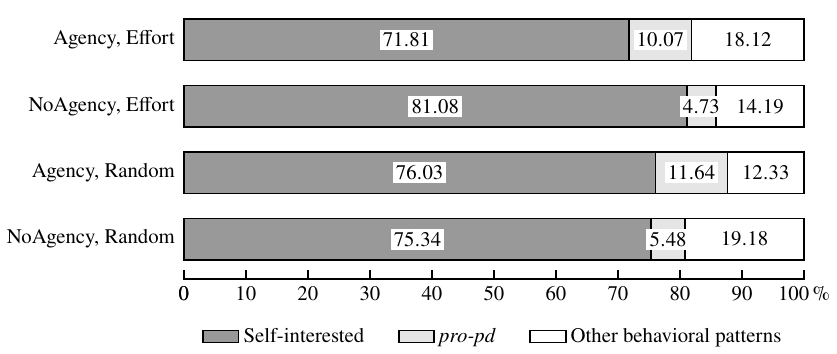}
    \caption{Behavioral patterns across the four conditions from Study 4, \emph{Approach}}
    \label{fig:overview_4b}
\end{figure}

The \textit{No Agency} treatments closely replicate our results from Study 3 (see Figure \ref{fig:overview_4b}).
The majority of consumers maximizes their income by staying in the same store for all four rounds and the price-discriminating store sells roughly the same amount before and after the introduction of discounts for low-income consumers (\textit{Effort:} $t = 0.29, p = 0.769$; \textit{Random:} $t = 0.27, p = 0.787$). 

Once consumers are endowed with some agency regarding future store prices, their purchasing behavior changes significantly. The share of subjects classified as \textit{pro-pd} roughly doubles with 10\,\% in \textit{Effort-Agency} and 12\,\% in \textit{Random-Agency}. In \textit{Effort-Agency}, the price discriminating store sells 0.09 goods on average over the first two rounds, and 0.30 goods over rounds three and four -- a significant increase ($t = 3.89, p < 0.001$). The results for \textit{Random-Agency} are similar (0.08 vs. 0.32; $t = 4.04, p < 0.001$) and there are no significant treatment differences between the two ($t = 0.25, p = 0.802$). The seller introducing BPD after the second round sells significantly more in \textit{Effort-Agency} than in \textit{Effort-No Agency} ($t = -2.94, p = 0.013$) and more in \textit{Random-Agency} than in \textit{Random-No Agency} ($t = -2.21, p = 0.028)$. Concurrently, the regression model reveals a significant interaction between BPD and the \textit{Agency} treatment dummy on store choice. The source of endowment does not predict switching.

\subsection*{Discussion}
The results show that a salient market mechanism, which endows consumers with agency over future prices for low-income consumers, more than doubles costly support of BPD, while aversions persist. This is consistent with altruistic preferences influencing switching towards the price discriminating store. Given that consumers are uncertain about the importance of their purchase in determining future prices -- as is the case for many real-life market contexts --, the effect is surprisingly large. On the other hand, high-income subjects confronted with relatively disadvantageous price discounts appear unfazed by the detrimental effects of their choices on inequality and economic welfare. This suggests that a consumer's role in the pricing environment mediates the influence of altruism on sellers' behavioral constraints. Those who can express altruism through voluntary, costly action are more amenable to support benevolent price discounts, whereas those deciding between inaction and disapproval largely disregard the outcome of low-income consumers.

\section{Study 5: Switching Costs and Warm Glow}\label{s5}

People often derive utility from ``giving'' to others, irrespective of the generated output. These warm glow effects are one main explanation why government transfers often only imperfectly crowd-out private contributions to public goods or charitable organizations. We argue that supporting benevolent price discrimination by giving up part of one's income is a manifestation of warm glow. Sellers depend on the reactions of other consumers when implementing price discounts for people with lower income. However, changing one's purchasing habits can be costly. For example, consumers might have to break certain routines, sustain longer distances, learn about and try new brands and products, create new accounts, or incur additional search costs. Depending on the consumer's preferences as well as different market factors like the purchase environment (analog vs. digital), competition or the availability of substitutes, these costs will differ substantially in the real world. For consumer constraints, such higher expenses should have either no, or a negative effect on switching -- depending on the severity of aversions (see utility function \ref{eq6}). For consumers with the option to \textit{support} BPD, however, this might change. That is because, in addition to the negative effect of losing money, consumers can ``feel good'' about themselves and their decision to give up, i.e. donate, money in order to assist the seller introducing price discounts that allow low-income consumers to access the market (see utility function \ref{eq5}). Consequently, the higher the switching costs, the better they may feel about their good deed. To test this theory, Study 5 varies switching costs for both \textit{Approach} and \textit{Avoid}. Following warm glow, we expect higher switching costs in \textit{Approach} to correlate with stronger support for BPD. For \textit{Avoid}, varying switching costs serves as a robustness check. Finally, we also provide a quick overview of two additional robustness checks that (1) vary the willingness to pay of low-income consumers and (2) use qualitative data to confirm the validity of our behavioral interpretations.

\subsection*{Experimental Design}
Since consumers in the prior \textit{Approach} studies did not condition their behavior on the origin of endowment differences, we choose to allocate incomes randomly. Furthermore, consumers have agency over future prices. We implement three treatments with switching costs of either 5, 15 or 30 Coins.

\subsubsection*{Results}

\begin{figure}[htbp]
\begin{center}
\begin{minipage}{0.9\linewidth}
\includegraphics[width=1\textwidth]{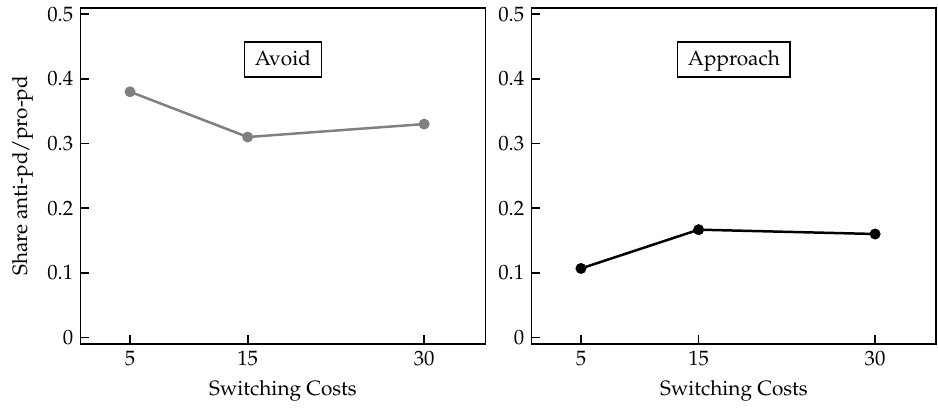}
\caption{Share of ``anti-pd'' and ``pro-pd'' subjects for different switching costs in \textit{Avoid} and \textit{Approach}.}
\label{fig:sc_overall}
\end{minipage}
\end{center}
\end{figure}

Figure \ref{fig:sc_overall} (right side) shows consumer ``pro-pd'' patterns for the three switching costs treatments. Compared to 5 Coins (10.7\%), the share of subjects classified as pro-pd increase under 15 (16.9\%) and 30 (16.3\%) Coins. BPD-stores in both conditions experience a significantly larger jump in sold goods after introducing price discounts (\textit{15}: 0.21 vs. 0.43, $t = 2.33, p = 0.02$; \textit{30}: 0.21 vs. 0.38, $t = 1.82, p = 0.07$), which is further supported by significant interaction effects in the panel regression (see Table \ref{tab:reg_approach}).

Like before, the majority of subjects follow a self-interested strategy and never switch stores (72\%/68\%/70\%) for switching costs of 5/15/30. Restricting the sample to participants who answered the control question correctly (76\%) does not change anything. Overall, these results support our conjunction that warm glow motivates costly support of BPD. While there are no differences between 15 and 30 Coins, switching towards the store with price discounts increases significantly compared to the baseline. Since costs ``c'' cause both positive and negative utility, it is not surprising that the effect levels off for larger amounts. 

\subsection*{5b: Switching Costs Avoid}

To assess the influence of switching costs on consumer aversions, we opt for the \textit{Avoid} paradigm where subjects earn their endowment through the effort task and have agency over the store's future prices. 

\subsubsection*{Results}

Results reveal moderate evidence for downward elasticity of consumer BPD aversions (see Figure \ref{fig:sc_overall} and Table \ref{tab:reg_avoid}). For switching costs of 5 Coins, the drop in average goods sold after the introduction of price discounts was significantly higher than for switching costs of 15 ($-0.91$ vs. $-0.70$; $t = -1.71, p = 0.045$) and qualitatively higher than for switching costs of 30 Coins ($-0.91$ vs. $-0.74$; $t = - 1.40, p = 0.082$). There was no difference between 15 and 30 Coins ($t = 0.30, p = 0.619$).\footnote{Restricting the sample to subjects who correctly answered the control question about consumer purchases affecting store prices in a future HIT (75\,\%) amplifies these results for switching costs of 15 (5: $-0.82$ vs. 15: $-0.51$; $t = -2.30, p = 0.011$), but not for switching costs of 30 (5: $-0.82$ vs. 30: $-0.65$; $t = -1.23, p = 0.110$).} Finally, the regression analysis reveals no significant treatment dummy (Table \ref{tab:reg_avoid}), confirming that consumer aversions towards BPD are relatively inelastic to changes in switching costs. 

\subsection*{Robustness Checks: Low-Income Consumer WTP and Consumer Reasoning}
We ensure the validity of our findings with two additional robustness checks that are detailed in the online appendix.\footnote{\url{https://osf.io/sztnh/?view_only=fc170dcbfeda431e994b23ed095d8b4b}} First, qualitative data from a post-experimental open-ended question in studies 2, 4 and 5 confirms (1) the validity of our behavioral labels and (2) high data quality. Consumer subjects consistently refer to unfair price discounts as their reason for switching away from the price-discriminating store, while transaction costs discipline many other consumers to act in their monetary self-interest. Second, we conduct a sixth large study (see Table \ref{tab:study_overview}) that varies low-income consumers' willingness to pay ($w_{lc} \in \{70, 100, 130\}$) such that high-income either experience (1) disadvantageous purchase-level inequity, (2) no purchase-level inequity or (3) advantageous purchase-level inequity. The data confirms the robustness of both \textit{Avoid} and \textit{Approach} towards relatively lower product valuations from low-income consumers, and thereby controls for high-income consumer disadvantageous inequity aversions as a behavioral explanation.

\section{General Discussion}\label{s:discussion}
This paper is the first to analyze consumer reactions towards benevolent price discounts that favor low-income consumers who otherwise cannot participate in the market. We show that a large share of consumers is willing to costly switch away from the price-discriminating store, despite experiencing no price increases themselves. Qualitative data confirms that this behavior is mainly driven by feeling unfairly passed over after the introduction of discounts. Simply observing that other people pay less for the same good is enough to motivate large consumer movements. In contrast to previous theories about the existence of price discounts, there is no evidence that inequity attitudes can adequately explain or mitigate consumer reactions. Even when consumers know that everybody is priced according to the same rule, and that differences in endowment are purely due to random chance, the main results hold. Furthermore, allowing for altruism by tying current purchasing choices to future prices does not meaningfully change consumer aversions. Our results therefore suggest that distributive fairness concerns do not play an important role in determining consumer fairness perceptions of relatively disadvantageous prices. This interesting results partially contrasts well-established findings from other economic contexts such as bargaining or redistribution through e.g. taxation \citep{alesina05,alesina2011preferences,camerer2011behavioral,engelmann2004inequality,dimick2016altruistic,klor2010social}.

Second, we find that costly \textit{support} for benevolent price discounts can be partially explained by consumers being impure altruists \citep{andreoni1990impure}. As hypothesized, costly switching towards a seller who introduces price discounts for excluded low-income consumers is behaviorally similar to the private provision of a public good, such as charitable giving \citep{ottoni2017people}. Without any information about price discrimination, or when purchases have no future consequences, almost no consumer is willing to costly switch away from their current store. However, once we allow for altruism, support more than doubles. What is more, increasing the costs of switching \textit{further} increases costly support of the price discriminating seller, providing evidence for warm glow effects. Like before, inequity attitudes do not appear to have meaningful explanatory power.

Overall, consumer aversions consistently exceed consumer support, and the seller introducing price discounts loses a sizeable share of high-income consumers. While the most prominent purchasing strategy is always a monetary self-interested one, qualitative and experimental data suggests that the existence of transaction costs prohibits many averse consumers from switching. Thus, the easier it is for consumers to find another seller, the higher the risk of potentially losing a disproportionate number of current consumers.

\subsubsection*{Implications}
The results highlight the importance of clearly distinguishing between different types of consumers. For a firm, there are marked differences between attracting novel customers who need to make an affirmative decision towards the seller, and not alienating or keeping current customers who, at best, express their contentment by foregoing a switching choice. The latter do not appear to judge top-down price discounts as a desirable public good, and therefore also do not exhibit sensitivity to relevant pro-social peer preferences. This is even true in situations where discounts are explicitly and exclusively targeted at people who, through sheer luck, are not able to participate in the market at all, suggesting very strong initial aversions towards price discrimination. Hence, while some researchers have suggested overt transparency as a firm's strategy to signal benevolence if price discrimination is based on consumer income \citep{Rotemberg.2011}, our results indicate that this is, by and large, not necessarily a good strategy for sellers. Rather, and similar to situations where consumers react to disadvantageous price discrimination \citep[e.g.][]{allender2021price, leibbrandt2020behavioral}, firms should generally expect a substantial drop in demand by current customers. Examples such as student discounts or the SNAP program are are outliers, and can probably not be explained by consumer inequity aversion or altruism. 

Our results are also in line with \citet{li2016behavior}, whereby firms that are expecting consumer fairness concerns reduce differential pricing, e.g., low poaching prices. This reduces inefficient switching and consumer dis-utility from perceived price unfairness, but hurts overall consumer surplus. Consumer reactions constrain price setting that increases market participation, consumer welfare, joint welfare, and equity. One particular implication is that seller monitoring through price transparency regulations and consumer agency does not necessarily imply utilitarian improvements. In fact, it may even have the opposite effect, by preventing many consumers from receiving better deals.

Finally, there are reasons to believe that price discrimination \textit{can} function as a re-distributive mechanism that increases market participation for low-income consumers while also increasing joint welfare. Consumers do exhibit impurely altruistic preferences when differential pricing is not a top-down decision by their current seller. Leveraging these social preferences may reveal hitherto under-explored mechanisms to attract new customers and simultaneously improve consumer surplus, e.g., through marketing, CSR communication or establishing a direct relationship between customer purchases and targeted price discounts. These instruments, however, will often require the utilization of either targeted information or obfuscation, in order to circumvent the behavioral constraints exerted by a seller's current customer base.

\bibliography{\bib}
\appendix

\input{Regs}

\end{document}

%% file: tab_overview.tex
\begin{table}[]
\caption{Study Overview -- \textit{Repeated} Paradigm}
\begin{adjustbox}{max width=\textwidth}
\begin{tabular}{lcccccccccccc}
\toprule
                & \multicolumn{2}{c}{Study 1} & \multicolumn{2}{c}{Study 2} & \multicolumn{2}{c}{Study 3} & \multicolumn{2}{c}{Study 4} & \multicolumn{2}{c}{Study 5} & \multicolumn{2}{c}{Study 6 (OA)} \\ \midrule
                & Avoid       & Approach      & Avoid       & Approach      & Avoid       & Approach      & Avoid       & Approach      & Avoid       & Approach       & Avoid       & Approach       \\ \midrule
Transparent BPD & X           & X             & X           & --             &  $\checkmark$           &    $\checkmark$               &      $\checkmark$           &      $\checkmark$             &  $\checkmark$               &      $\checkmark$             &      $\checkmark$            &    $\checkmark$            \\[3pt]
No Intention    &      X       &        X       &   $\checkmark$               &     --              &     X        &       X        &         X    &         X      &         X    &          X      &        X     &        X        \\[3pt]
Effort          &      X       &          X     &        X     &       --          &    $\checkmark$             &    $\checkmark$               &    $\checkmark$            &    $\checkmark$               &    $\checkmark$            &     X           &      $\checkmark$           &  $\checkmark$                  \\[3pt]
Random          &    X         &        X       &        X     &       --      &     $\checkmark$            &   $\checkmark$                &    $\checkmark$             &    $\checkmark$               &     X        &     $\checkmark$               &      $\checkmark$           &    $\checkmark$                \\[3pt]
No Agency       &   $\checkmark$              &     $\checkmark$              &   $\checkmark$              &    --               &   $\checkmark$              &    $\checkmark$               &     $\checkmark$            &      $\checkmark$             &      X         &         X          &     X        &       X         \\[3pt]
Agency          &     X        &        X       &     X        &        -       &       X      &        X       &      $\checkmark$         &      $\checkmark$           &     $\checkmark$          &      $\checkmark$            &       $\checkmark$        &     $\checkmark$             \\[3pt]
Switching Costs &     10        &   10             &      10        &        --        &    10          &      10          &      10        &        10        &      5, 15, 30       &  5, 15, 30              &    10         &          10      \\[3pt]
$w_{\text{low\_income}}$ &     150        &   150             &      150        &        150        &    150          &      150          &      150        &        150 &     150       & 150      &    70, 100, 130         &  70, 100, 130    \\
\midrule
N               & 96 & 97       & 391 & --       & 197 & 200     & 586 & 589        & 367 & 444        & 479 & 505     \\
Female \%       & 47 & 47       & 52 & --       & 47 & 45        & 49 & 50        & 54 &  43     & 45 & 48 \\    
\bottomrule
\end{tabular}
\end{adjustbox}
\label{tab:study_overview}
\end{table}

%% file: Regs.tex

\begin{table}[htbp]
    \centering
     \caption{Panel logistic regression using random effects for \emph{Avoid}}

    \begin{tabularx}{\textwidth}{X*{5}{l}}
        \toprule
        & {Study 1} & {Study 2} & {Study 3} & {Study 4a} & {Study 5b} \\
        \cmidrule{2-6}
        & \multicolumn{1}{c}{\makecell{Coef.\\(se)}} & \multicolumn{1}{c}{\makecell{Coef.\\(se)}} & \multicolumn{1}{c}{\makecell{Coef.\\(se)}} & \multicolumn{1}{c}{\makecell{Coef.\\(se)}} & \multicolumn{1}{c}{\makecell{Coef.\\(se)}} \\
        \midrule
					PD        &   -5.353{***}&   -3.935***&   -4.700***&   -3.994***&   -5.273{***}\\
          &  (1.206)   &  (0.459)   &  (0.806)   &  (0.587)   &  (0.726)  \\
PD $\times$ \emph{Intention}&            &   -1.543** &            &            &         \\
          &            &  (0.617)   &            &            &       \\
PD $\times$ \emph{Effort}&            &            &    0.507   &   -1.300*  &   \\
          &            &            &  (0.852)   &  (0.754)   &    \\
PD $\times$ \emph{Agency}&            &            &            &    0.380   &          \\
          &            &            &            &  (0.739)   &      \\
PD $\times$ \emph{Effort} $\times$ \emph{Agency}&            &            &         &    0.351  & \\
          &            &            &            &  (1.075)   &     \\
PD $\times$ \emph{15 Coins}&            &            &            &            &     0.598 \\
          &            &            &            &            &         (0.849) \\
PD $\times$ \emph{30 Coins}&            &            &            &            &    0.732  \\
          &            &            &            &            &          (0.812)   \\
\emph{Intention} &            &    0.969*  &            &            &            \\
          &            &  (0.520)   &            &            &         \\
\emph{Effort}    &            &            &   -0.412   &    1.119   &    \\
          &            &            &  (0.731)   &  (0.683)   &   \\
\emph{Agency}    &            &            &            &    0.270   &         \\
          &            &            &            &  (0.646)   &        \\
\emph{Effort} $\times$ \emph{Agency}&            &            &            &   -1.050   & \\
          &            &            &            &  (0.951)   &       \\
\emph{15 Coins}  &            &            &            &            &            0.504 \\
          &            &            &            &            &          (0.731) \\
\emph{30 Coins}  &            &            &            &            &           0.096  \\
          &            &            &            &            &         (0.683) \\
SCO       &    0.513*  &   -0.237   &   -0.086   &   -0.375***&    -0.117      \\
          &  (0.304)   &  (0.172)   &  (0.199)   &  (0.144)   &  (0.165)    \\
Constant  &    3.383** &    5.582***&    5.647***&    6.907***&   5.758*** \\
          &  (1.584)   &  (0.965)   &  (1.263)   &  (0.958)   &   (0.992)    \\
\midrule
N         &\multicolumn{1}{c}{384}   &\multicolumn{1}{c}{1564}   &\multicolumn{1}{c}{788}   &\multicolumn{1}{c}{2344}   &\multicolumn{1}{c}{1468}    \\
AIC       &\multicolumn{1}{c}{292.194}   &\multicolumn{1}{c}{1267.133}   &\multicolumn{1}{c}{634.181}   &\multicolumn{1}{c}{1734.051}   &\multicolumn{1}{c}{1129.807}   \\
BIC       &\multicolumn{1}{c}{307.996}   &\multicolumn{1}{c}{1299.263}   &\multicolumn{1}{c}{662.198}   &\multicolumn{1}{c}{1791.647}   &\multicolumn{1}{c}{1172.140}  \\
 					\bottomrule
   \end{tabularx}
    \label{tab:reg_avoid}
    \note{Table reports results of panel logistic regressions using random effects and a cluster–robust VCE estimator. Dependent variable is a dummy variable that equals 1 if participant bought a good in the price discriminating store, 0 otherwise. Independent variables: "PD" equals 1 in Round 2 and 3, 0 otherwise; \emph{Intention}, \emph{Effort}, \emph{Agency}, \emph{15 ECU} and \emph{30 ECU} are dummy variables for the experimental conditions; Reference groups are omitted from the table; "SCO" is mean of Social comparison orientation scale by \citet{gibbons99}.  --- \signstars}
\end{table}

\begin{table}[htbp]
    \centering
     \caption{Panel logistic regression using random effects for \emph{Approach}}
    \begin{tabularx}{\textwidth}{X*{4}{l}}
        \toprule
        & {Study 1} & {Study 3} & {Study 4b} & {Study 5a}  \\
        \cmidrule{2-5}
        & \multicolumn{1}{c}{\makecell{Coef.\\(se)}} & \multicolumn{1}{c}{\makecell{Coef.\\(se)}} & \multicolumn{1}{c}{\makecell{Coef.\\(se)}} & \multicolumn{1}{c}{\makecell{Coef.\\(se)}} \\
        \midrule
					PD        &    0.636   &    0.409   &    0.120   &  1.667*** \\
          &  (0.666)   &  (0.586)   &  (0.437)  & (0.476)  \\
PD $\times$ \emph{Effort}&            &    0.184   &    0.026  &  \\
          &            &  (0.774)   &  (0.654)   &  \\
PD $\times$ \emph{Agency}&            &            &    1.923***&  \\
          &            &            &  (0.650)   \\
PD $\times$ \emph{Effort} $\times$ \emph{Agency}&            &            &   -0.258  & \\
          &            &            &  (0.913)   &   \\
PD $\times$ \emph{15 Coins} &            &            &    & 1.694**   \\
          &            &            &    &  (0.665)  \\
PD $\times$ \emph{30 Coins} &            &            &   & 1.500**  \\
          &            &            &    & (0.726)  \\
          
\emph{Effort}    &            &   -0.012   &   -0.384   &   \\
          &            &  (0.568)   &  (0.448)   &  \\
\emph{Agency}    &            &            &   -0.992** &  \\
          &            &            &  (0.503)   &    \\
\emph{Effort} $\times$ \emph{Agency}&            &            &    0.527   &     \\
          &            &            &  (0.709)   \\
          
\emph{15 Coins}    &            &     &      &  \\
          &            &   &     & (0.590) \\
\emph{30 Coins}    &            &            &    & -0.955  \\
          &            &            &   &    (0.637)  \\
          
SCO       &   -0.308   &   -0.076   &   -0.064   & -0.093   \\
          &  (0.239)   &  (0.193)   &  (0.121)   & (0.133)   \\
Constant  &   -2.731** &   -3.709***&   -3.420***&  -4.399*** \\
          &  (1.229)   &  (1.037)   &  (0.675)   &  (0.721)    \\
\midrule
N         &\multicolumn{1}{c}{388}   &\multicolumn{1}{c}{800}   &\multicolumn{1}{c}{2356}  &\multicolumn{1}{c}{1776}  \\
AIC       &\multicolumn{1}{c}{153.150}   &\multicolumn{1}{c}{381.708}   &\multicolumn{1}{c}{1246.513}  & \multicolumn{1}{c}{1132.023}   \\
BIC       &\multicolumn{1}{c}{168.994}   &\multicolumn{1}{c}{409.816}   &\multicolumn{1}{c}{1304.160}  & \multicolumn{1}{c}{1175.880}   \\
\bottomrule
   \end{tabularx}
    \label{tab:reg_approach}
    \note{ Table reports results of panel logistic regressions using random effects and a cluster–robust VCE estimator. Dependent variable is a dummy variable that equals 1 if participant bought a good in the price discriminating store, 0 otherwise --- Independent variables: ``PD'' equals 1 in Round 2 and 3, 0 otherwise; \emph{Effort} and \emph{Agency}  are dummy variables for the experimental conditions; Reference groups are omitted from the table; ``SCO'' is mean of Social comparison orientation scale by \citet{gibbons99} --- \signstars}
\end{table}